\documentclass{PoS}

\newcommand{\id}{{\bf 1}}
\newcommand{\Tr}{\mathrm{Tr}\,}

\title{
Index theorem and overlap formalism with naive and minimally doubled
fermions\thanks{RIKEN-MP-34, YITP-11-87}
}

\ShortTitle{Index theorem and overlap formalism}

\author{\speaker{Taro Kimura}\\%\thanks{A footnote may follow.}\\
        Department of Basic Science, University of Tokyo\\
	Mathematical Physics Laboratory, RIKEN\\
        E-mail: \email{kimura@dice.c.u-tokyo.ac.jp}}

\author{Michael Creutz\thanks{Authored under contract number
DE-AC02-98CH10886 with the U.S.~Department of Energy.  Accordingly,
the U.S. Government retains a non-exclusive, royalty-free license to
publish or reproduce the published form of this contribution, or allow
others to do so, for U.S.~Government purposes.}\\
        Brookhaven National Laboratory\\
        E-mail: \email{creutz@bnl.gov}}

\author{Tatsuhiro Misumi\\
        Yukawa Institute for Theoretical Physics, Kyoto University\\
        E-mail: \email{misumi@yukawa.kyoto-u.ac.jp}}

\abstract{
We present a theoretical foundation for the index theorem in naive and
minimally doubled lattice fermions by studying the spectral flow of a
Hermitean version of Dirac operators. We utilize the point splitting method 
to implement flavored mass terms, which play an important role in constructing 
proper Hermitean operators. We show the spectral flow correctly detects the 
index of the would-be zero modes which is determined by gauge field topology 
and the number of species doublers. Using the flavored mass terms, we present 
new types of overlap fermions from the naive fermion kernels, with a number of 
flavors that depends on the choice of the mass terms. 
}

\FullConference{ The XXIX International Symposium on Lattice Field Theory - Lattice 2011\\
July 10-16, 2011\\
Squaw Valley, Lake Tahoe, California}

\begin{document}

\section{Introduction}\label{sec:introduction}

%The index is hidden in lattice fermions with species doublers since the
%index effect cancels between doubling pairs.
Ref.~\cite{Adams1} recently presented how to identify the would-be zero
modes and their chiralities with staggered fermions, and its application
to the staggered versions of the overlap and the Wilson fermions.
There have been some relating works on this topic so far, e.g. the 
single taste generalization \cite{Hoel}, the numerical efficiency \cite{FKP} 
of the staggered overlap and the phase structure and the chiral limit of the
staggered Wilson fermion \cite{CKM2}.

This approach is applied by the present authors to other fermions
with species doublers, i.e. naive and minimally doubled fermions
in Ref.~\cite{CKM1}, which this report is based on. We identify species 
in naive and minimally doubled fermions as flavors by using the 
point-splitting method \cite{Creutz}, and then define proper flavored-mass 
terms to extract the index in the spectral flow. We also present new versions 
of overlap fermions built on these fermion kernels. Especially we construct a 
single-flavor naive overlap fermion by using a certain flavored-mass term which 
assigns negative mass to only one of species.

\section{Point-splitting and flavored-mass terms}\label{sec:point}

We first introduce the point-splitting method to obtain flavored-mass 
terms for minimally doubled fermions. In this paper we concentrate on 
the Karsten-Wilczek fermion \cite{KW} rather than other formulations, 
i.e. Bori\c{c}i-Creutz fermion \cite{BC} and so on (see \cite{MCK} for review).
We consider the following Dirac operator for the $d=4$ minimally doubled
fermion in the momentum space,
\begin{equation}
 D_{\rm md}(p) = 
  i \sum_{k=1}^3 \gamma_k \sin p_k + \frac{i\gamma_4}{\sin \alpha}
  \left(
   \cos \alpha + 3 - \sum_{\mu=1}^4 \cos p_\mu
  \right),
\end{equation}
with a free parameter $\alpha$.
It has only two zeros located at $p=(0,0,0,\pm \alpha)$.
These two species are not equivalent since the gamma matrices are
differently defined between them as $\gamma_\mu' = \Gamma^\dag
\gamma_\mu \Gamma$. In this case the transformation matrix is given 
by $\Gamma=i\gamma_4\gamma_5$.

\begin{figure}[t]
 \begin{center}
  \includegraphics[width=13em]{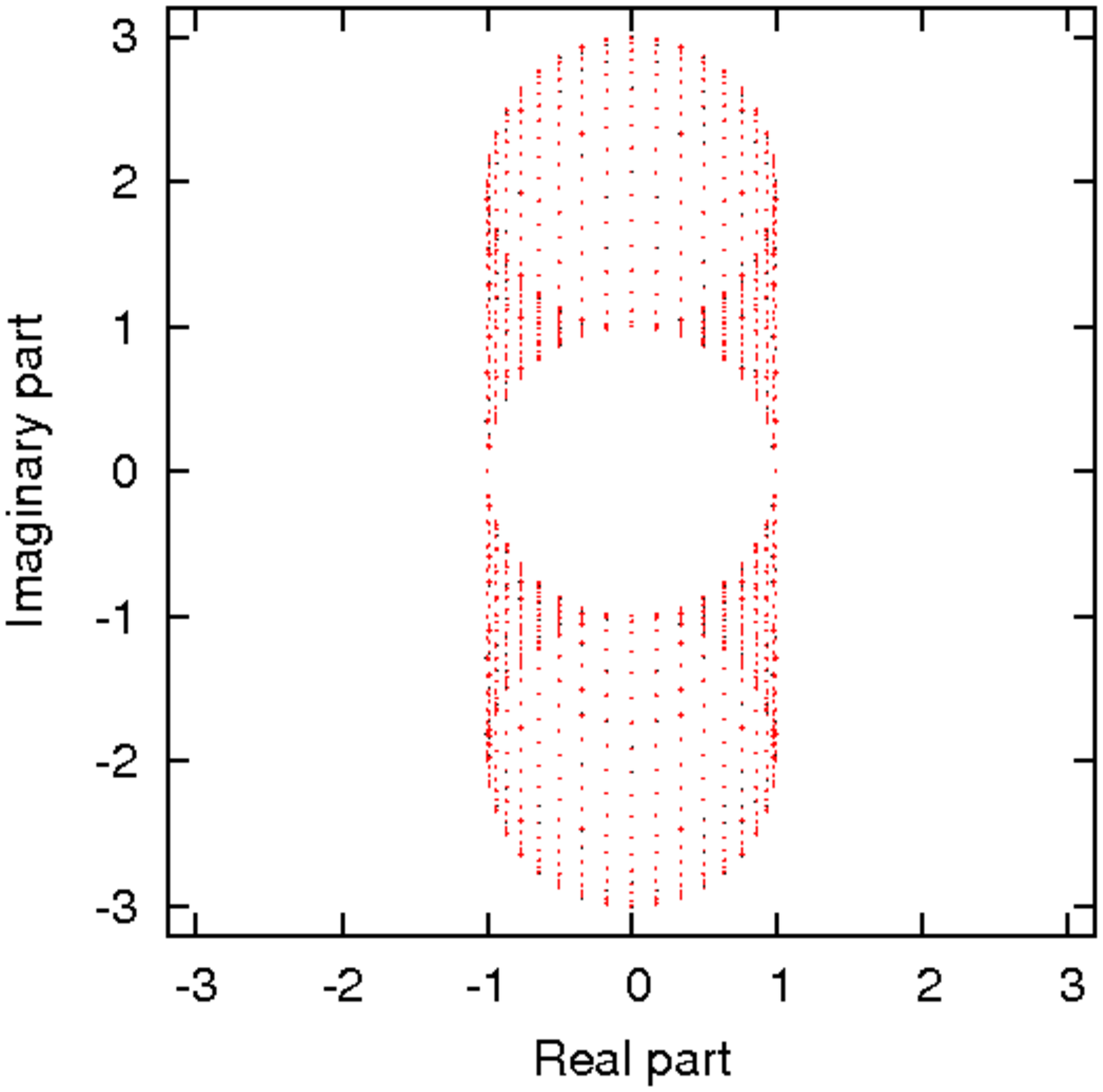} \qquad
  \includegraphics[width=13em]{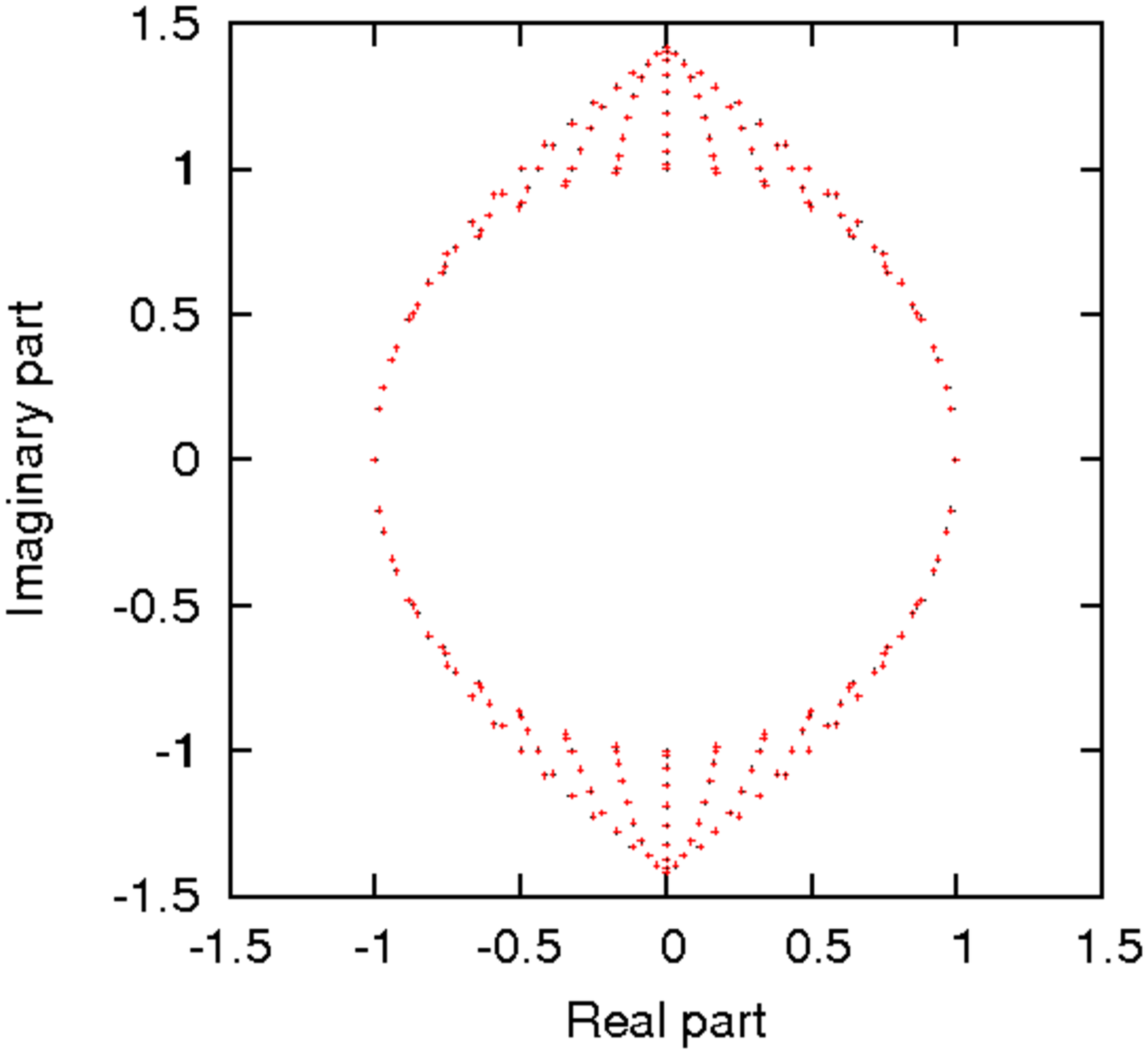}
  \vspace{-1.5em}
 \end{center}
 \caption{Dirac spectra with flavored mass terms for 2-dimensional
 free theory: (left) minimally doubled and (right) naive fermions with a
 mass parameter $m=1$.} 
 \label{Dirac_spec}
\end{figure}

The point splitting identifies these inequivalent species as independent
flavors \cite{Creutz}. In this method each flavor field is defined so that the 
associated fermion propagator includes only a single pole by multiplying 
factors removing the other pole,
\begin{equation}
 u(p-\alpha e_4) = \frac{1}{2} 
  \left( 1 + \frac{\sin p_4}{\sin \alpha} \right) \psi(p), \quad
 d(p+\alpha e_4) = \frac{1}{2} \Gamma
  \left( 1 - \frac{\sin p_4}{\sin \alpha} \right) \psi(p) .
\end{equation}
Here we work in the momentum space, but the forms in the position 
space are also obtained through the Fourier transformation even when 
the gauge fields are turned on \cite{Creutz}. To regard the two fields as 
flavors we consider a flavor-multiplet 
field as following,
\begin{equation}
 \Psi(p) = 
  \left(
   \begin{array}{c}
    u(p-\alpha e_4) \\ d(p+\alpha e_4) 
   \end{array}
  \right),
\end{equation}
where $e_\mu$ stands for the reciprocal vector. We note in this representation of 
the fermion field $\gamma_5$ multiplication is expressed as
\begin{equation}
 \gamma_5 \psi(p) \quad \longrightarrow \quad
  \left(
   \begin{array}{cc}
    +\gamma_5 & \\ 
    & -\gamma_5 \\
   \end{array}
  \right) \Psi(p)
  = (\gamma_5 \otimes \tau_3) \Psi(p) .
\end{equation}
This shows the usual $\gamma_5$ is a flavor non-singlet gamma-5 as 
$\gamma_5 \otimes \tau_3$ in the doubler multiplet. Now our purpose 
is to find a flavored-mass term to assign different masses to the two species.
For this purpose we introduce the flavor structure $\tau_{3}$ in the mass term 
and rewrite it in the usual representation of the fermion field as
\begin{equation}
 \bar\Psi(p) ( \id \otimes \tau_3 ) \Psi(p)
  = \bar u u - \bar d d 
  = \frac{\sin p_4}{\sin \alpha} \bar \psi(p) \psi(p).
\end{equation}
In the position space this flavored-mass term is given by 
\begin{equation}
 M_{\tau_3} = \frac{-im_{\tau_3}}{2\sin \alpha} (T_{+4} - T_{-4})
  \equiv M_{\rm md}, 
  \label{flv_mass_md}
\end{equation}
with the translation operator $T_{\pm \mu} \psi_x =U_{x,x\pm\mu}\psi_{x\pm\mu}$.
Although this flavored-mass term satisfies the gamma-5 hermiticity,
in the non-hermitian operator $D_{\rm md} - M_{\rm md}$ 
the kinetic and flavored-mass terms do not commute 
$[D_{\rm md}, M_{\rm md}]\not=0$ in the presence of the link variable.
It results in the complex-valued Dirac spectrum, which is split into two branches
in this case. It indicates the flavored-mass term assigns one of species positive mass 
and the other negative mass. 
In Fig.~\ref{Dirac_spec}~(left) we show a numerical result of complex
eigenvalues of the Dirac operator for the $d=2$ free case with a
parameter being $\alpha=\pi/2$.

We then investigate the doubler-multiplet for the naive lattice fermion and 
the corresponding flavored-mass terms. The Dirac operator for the 
naive fermion in general dimensions is given by
\begin{equation}
 D_{\rm naive}(p) = i \sum_{\mu=1}^d \gamma_\mu \sin p_\mu.
\end{equation}
For simplicity we here consider the $d=2$ case.
It has four zeros, thus we introduce four associated point-splitting fields as
\begin{eqnarray}
 \psi_{(1)}(p-p_{(1)}) & = & \frac{1}{4} 
  (1+\cos p_1)(1+\cos p_2) \Gamma_{(1)} \psi(p), \\
 \psi_{(2)}(p-p_{(2)}) & = & \frac{1}{4} 
  (1-\cos p_1)(1+\cos p_2) \Gamma_{(2)} \psi(p), \\
 \psi_{(3)}(p-p_{(3)}) & = & \frac{1}{4} 
  (1+\cos p_1)(1-\cos p_2) \Gamma_{(3)} \psi(p), \\
 \psi_{(4)}(p-p_{(4)}) & = & \frac{1}{4} 
  (1-\cos p_1)(1-\cos p_2) \Gamma_{(4)} \psi(p),
\end{eqnarray}
where $p_{i} (i=1,2,3,4)$ stand for the zeros of four species.
Their locations of zeros, chiral charges and transformation matrices
for the sets of gamma matrices $\gamma_{(i)}^\mu =
\Gamma^\dag_{(i)} \gamma^\mu \Gamma_{(i)}$, are listed in
Table~\ref{zeros_list}.

\begin{table}[t]
 \begin{center}
  \begin{tabular}{cccc} \hline \hline
   label & position & $\chi$ charge & $\Gamma$ \\ \hline
   $1$ & $(0,0)$ & $+$ & $\id$ \\ 
   $2$ & $(\pi,0)$ & $-$ & $i\gamma_1 \gamma_5$ \\ 
   $3$ & $(0,\pi)$ & $-$ & $i\gamma_2 \gamma_5$ \\ 
   $4$ & $(\pi,\pi)$ & $+$ & $\gamma_5$ \\ \hline \hline
  \end{tabular}
 \end{center}
 \caption{Chiral charges and transformation matrices for each of zeros
 in the $d=2$ naive fermions with $\gamma_1 = \sigma_1$, $\gamma_2 =
 \sigma_2$ and $\gamma_5 = \sigma_3$.}
 \label{zeros_list}
\end{table}

The doubler-multiplet field is given by
\begin{equation}
 \Psi(p) =
  \left(
   \begin{array}{c}
    \psi_{(1)} (p-p_{(1)}) \\
    \psi_{(2)} (p-p_{(2)}) \\
    \psi_{(3)} (p-p_{(3)}) \\
    \psi_{(4)} (p-p_{(4)}) \\
   \end{array}
  \right) ,
\end{equation}
and in this case the $\gamma_5$ multiplicatioin is expressed as
\begin{equation}
 \gamma_5 \psi(p) \quad \longrightarrow \quad
  \left(
   \begin{array}{cccc}
    +\gamma_5 & & & \\
    & -\gamma_5 & & \\
    & & -\gamma_5 & \\
    & & & +\gamma_5 \\
   \end{array}
  \right) \Psi(p)
  = (\gamma_5\otimes(\tau_3\otimes\tau_3)) \Psi(p),
\end{equation}
where we express the 4-flavor structure in the doubler-multiplet
by two direct products of the Pauli matrix.
For our purpose of obtaining the flavored-mass terms to split species, 
we introduce the following flavor structure in the mass term this time.
\begin{equation}
 \bar\Psi(p) (\id\otimes(\tau_3\otimes\tau_3)) \Psi(p)
  = \cos p_1 \cos p_2 \bar \psi(p) \psi(p).
\end{equation}
This flavor structure gives two positive and two negative eigenvalues, which
implies there are two species with positive and the others with negative mass. 
Its position-space expression is given by
\begin{eqnarray}
  M_{\tau_3\otimes\tau_3} & = & m_{\tau_3\otimes\tau_3} \sum_{sym.} C_1 C_2
  \equiv M_{\rm naive},
  \label{flv_mass_n1} 
\end{eqnarray}
where we define $C_\mu = (T_{+\mu} + T_{-\mu})/2$ and ($\sum_{sym.}$)
stands for symmetric summation over the order of the factors.
In Fig.~\ref{Dirac_spec}~(right) we show the eigenvalues of the Dirac
operator with the flavored mass term, $D_{\rm naive} - M_{\rm naive}$.
The spectrum is split into two doubled branches crossing the real axis
at $|m|$, which indicates the flavored-mass term assigns 
positive mass to two species while negative mass to the other two.

Furthermore we can consider other kinds of flavored mass terms,
\begin{eqnarray}
 M_{\tau_3\otimes\id} & = & \frac{1}{2} m_{\tau_3\otimes\id}
  \sum_{sym.} (1+C_1^2)C_2
  \simeq m_{\tau_3\otimes\id} C_2,
  \label{flv_mass_n2} \\
 M_{\id\otimes\tau_3} & = & \frac{1}{2} m_{\id\otimes\tau_3}
  \sum_{sym.} C_1 (1+C_2^2)
  \simeq m_{\id\otimes\tau_3} C_1 .
  \label{flv_mass_n3}
% M_{\id\otimes\id} & = & m_{\id\otimes\id}
%  \label{flv_mass_n4}
\end{eqnarray}
These terms are utilized to discuss the overlap formalism in
section~\ref{sec:overlap}.
In the $d=4$ case wa can apply the same approach to obtain a proper
flavored mass terms \cite{CKM1}.

\section{Spectral flow and the index theorem}\label{sec:spectral}

In the continuum field theory the index is defined as the difference
between the numbers of zero modes of the Dirac operator with positive
and negative chirality, $n_+$ and $n_-$.
The statement of the index theorem is that the index is just equal to a
topological charge $Q$ of a background gauge configuration up to a sign
factor depending on its dimensionality,
\begin{equation}
 n_+ - n_- = (-1)^{d/2} Q .
\end{equation}
To detect the index of the Dirac operator, it is useful to introduce a
certain Hermitean version of the Dirac operator $H(m) = \gamma_5(D-m)$,
%\begin{equation}
% H(m) = \gamma_5(D-m),
%\end{equation}
where zero modes of the Dirac operator with $\pm$ chirality
correspond to eigenmodes of this Hermitean operator with
eigenvalues $\lambda(m)=\mp m$. 
If we consider the flow of the eigenvalues $\lambda(m)$ as the mass
varies, those corresponding to zero modes will cross the origin with
slopes $\mp 1$ depending on their chirality. 
The non-zero eigenmodes of $D$, in contrast, occur in pairs which are
mixed by $H$ and cannot cross zero. 
Therefore the index of the Dirac operator is given by minus the spectral
flow of the Hermitean operator, which stands for the net number of
eigenvalues crossing the origin, counted with sign depending on the
slope.

\begin{figure}[t]
 \begin{center}
  \vspace{-4em}
  \includegraphics[width=13em,angle=-90]{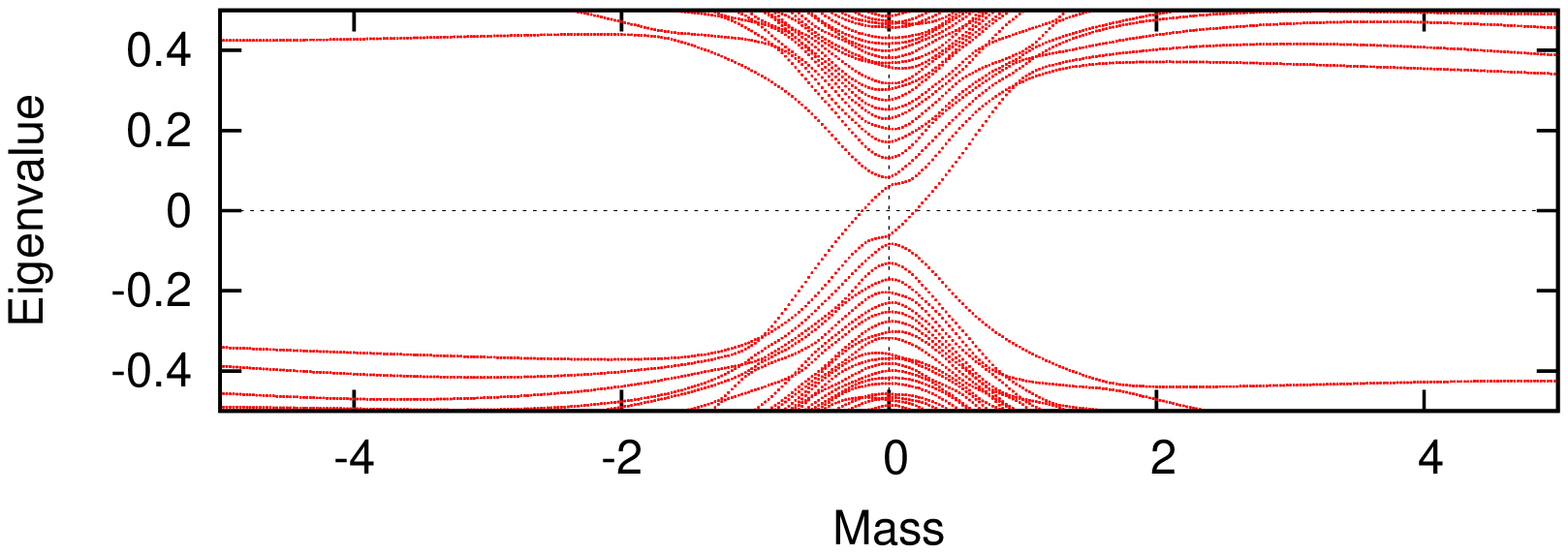} 
  \includegraphics[width=13em,angle=-90]{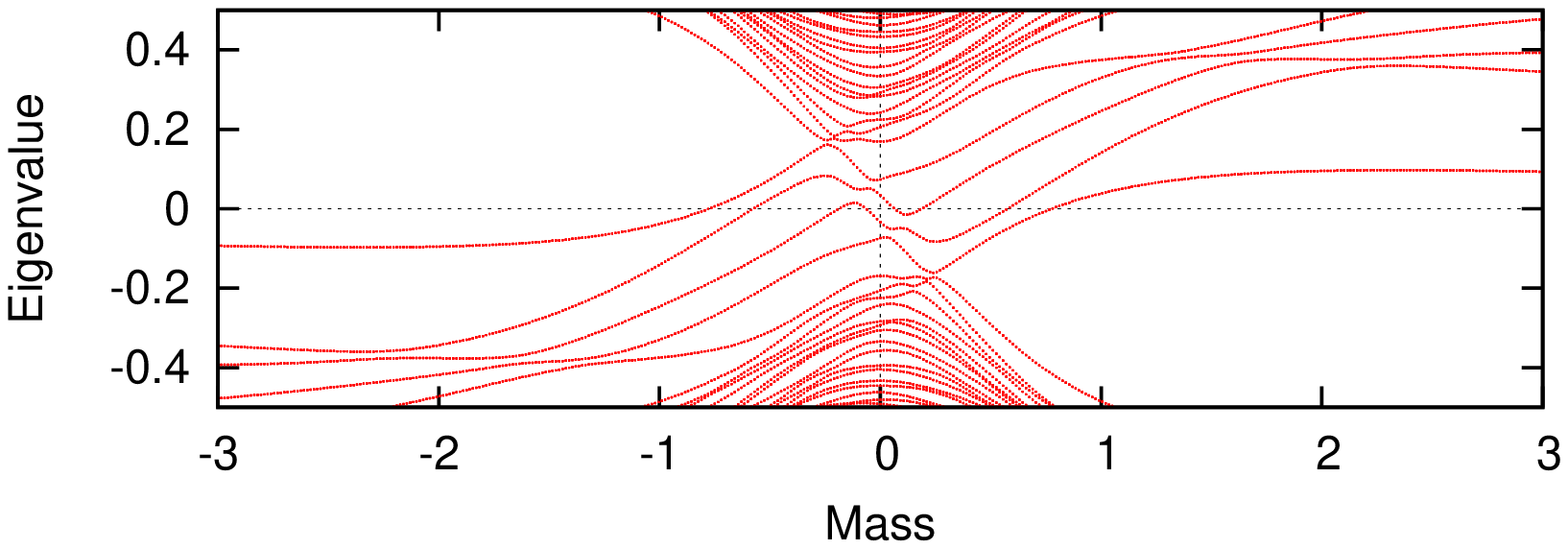}
  \vspace{-4em}
 \end{center}
 \caption{Spectral flows of the Hermitean operator based on the naive
 fermion for the 2-dimensional case with a topological charge (left) $Q=1$ and (right) $Q=2$.}
 \label{sf_naive}
\end{figure}

The index theorem for lattice fermions is also obtained from the spectral
flow for not only the Wilson \cite{ind_wilson}, but also the fermions with
species doublers \cite{Adams1,CKM1}.
However In the latter cases, the index apparently cancels between pairs of
doublers, thus an eigenvalue flow with a simple mass term does not
properly capture gauge field topology.
This difficulty is resolved by introducing proper flavored-mass
terms for staggered \cite{Adams1}, minimally doubled and naive fermions
\cite{CKM1}. The Hermitean versions of naive and minimally-doubled Dirac 
operators with flavored-mass terms are
\begin{equation}
 H_{\rm md}(m) = \gamma_5 (D_{\rm md} - M_{\rm md}), \qquad
 H_{\rm naive}(m) = \gamma_5 (D_{\rm naive} - M_{\rm naive}), 
 \label{Hermitean_ops}
\end{equation}
where $M_{\rm md}$ and $M_{\rm naive}$ are the flavored-mass terms
in (\ref{flv_mass_md}) and (\ref{flv_mass_n1}), and $m$ stands for
$m_{\tau_{3}}$ for the minimally doubled fermion and 
$m_{\tau_{3}\otimes\tau_{3}}$ for the naive fermion.
Note that the mass terms of these Hermitean operators are
flavor-singlet as $\gamma_5 M_{\rm md} \simeq \gamma_5 \otimes \id$, 
$\gamma_5 M_{\rm naive} \simeq \gamma_5 \otimes \id$, which 
is essential in detecting the correct index.
Fig.~\ref{sf_naive} shows the numerical result of the spectral flow for
the $d=2$ naive fermion with a topological charge $Q=1$ and $Q=2$,
respectively.\footnote{See \cite{CKM1} for details of the numerical
simulation and the case with minimally doubled fermion.}
There are doubled crossings around the origin, and the number of crossings counted 
depending on slopes should be the index related to the topological charge.
Taking account of the sign of the slope of the crossings, 
these results satisfies the index theorem for the naive fermion given by
\begin{equation}
% {\rm Index}(D_{\rm md}) = 2 (-1)^{d/2} Q, \qquad
 {\rm Index}(D_{\rm naive}) = 2^d (-1)^{d/2} Q ,
\end{equation}
where the factor reflects $2^d$ species.
We can study minimally doubled
fermions in a parallel way  \cite{CKM1}.

\section{Overlap formalism}\label{sec:overlap}

The establishment of the index theorem, as discussed in
section~\ref{sec:spectral}, leads to new versions of overlap fermions 
\cite{Overlap} based on naive and minimally doubled fermions,
\begin{equation}
 D_{\rm ov} = 1 + \gamma_5 \frac{H(m)}{\sqrt{H(m)^2}}
\end{equation}
where we now substitute the Hermitean operators defined in (\ref{Hermitean_ops}).
Here $m$ stands for $m_{\tau_{3}}$ or $m_{\tau_{3}\otimes\tau_{3}}$ again.
We note it satisfies the Ginsparg-Wilson relation 
$\left\{\gamma_5, D_{\rm ov}\right\} = D_{\rm ov} \gamma_5 D_{\rm ov}$.
We can obtain the index of the overlap fermion from this Ginsparg-Wilson
relation
\begin{equation}
 {\rm Index}(D_{\rm ov}) = - \frac{1}{2}
  \Tr \left(\frac{H(m)}{\sqrt{H(m)^2}}\right)
  = - \frac{1}{2} \Tr \, \mathrm{sgn} \, H(m) .
\end{equation}
Note that the half of the species which have negative mass
are converted into physical massless modes in the overlap formulation
while the others with positive mass become massive and decouple in the
continuum limit. 
This is because the flavored-mass terms we introduce for minimally doubled 
and naive fermions assign negative mass to half of species and positive mass 
to the others. This reduction of flavors also affects the indices of the 
Dirac operators,
\begin{equation}
 {\rm Index}(D_{\mbox{\scriptsize md-ov}}) 
  = \frac{1}{2} {\rm Index}(D_{\rm md}), \qquad
 {\rm Index}(D_{\mbox{\scriptsize n-ov}}) 
 = \frac{1}{2} {\rm Index}(D_{\rm naive}) .
\end{equation}
This relation relies on the property of the Hermitean operator $\gamma_5
H(m) \gamma_5 = -H(-m)$.
This is also the case with the staggered overlap, 
but not the case with the usual Wilson fermion.

\begin{figure}[t]
 \begin{center}
  \vspace{-3em}
  \includegraphics[width=13em,angle=-90]{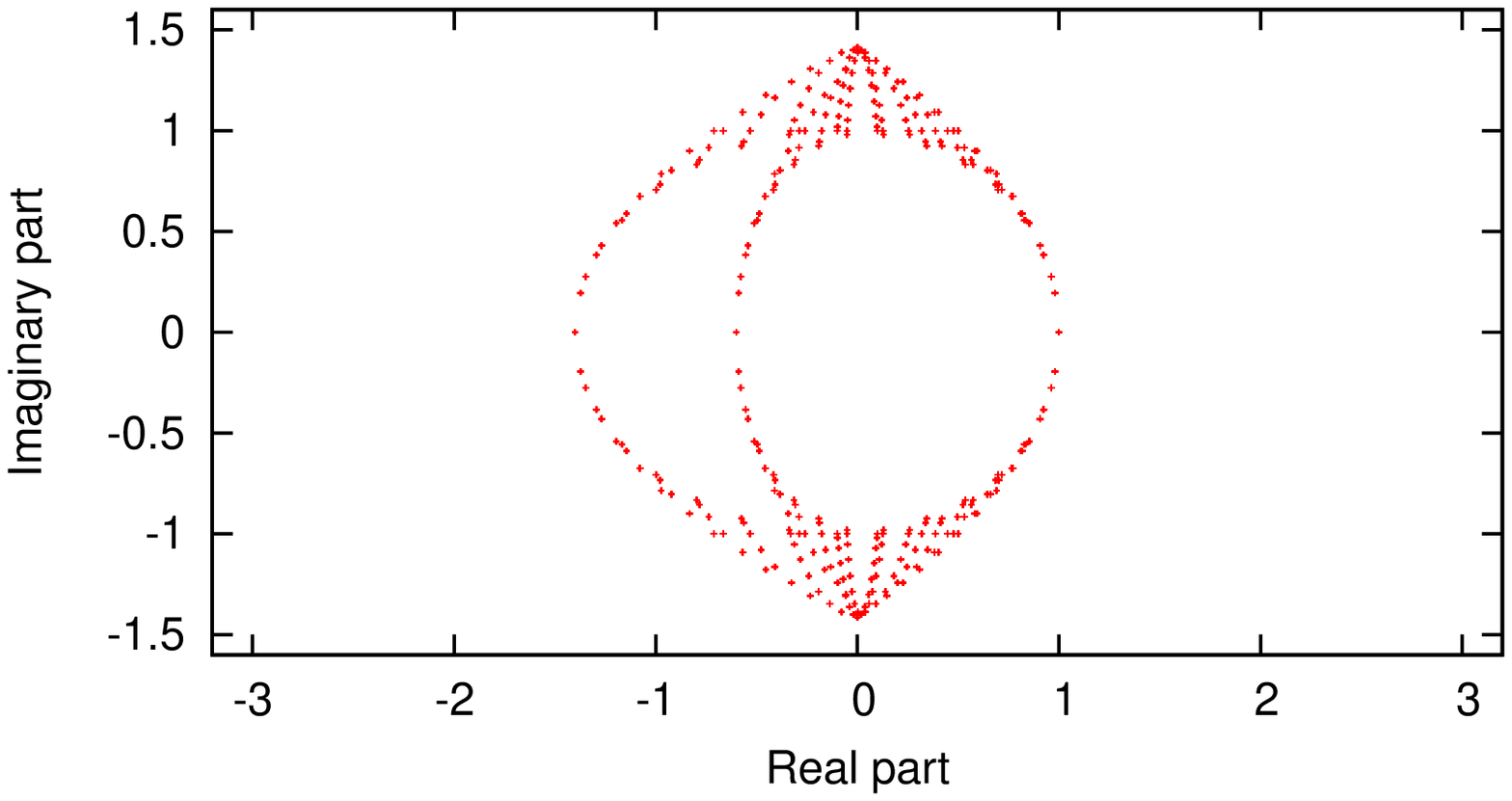}
  \includegraphics[width=13em,angle=-90]{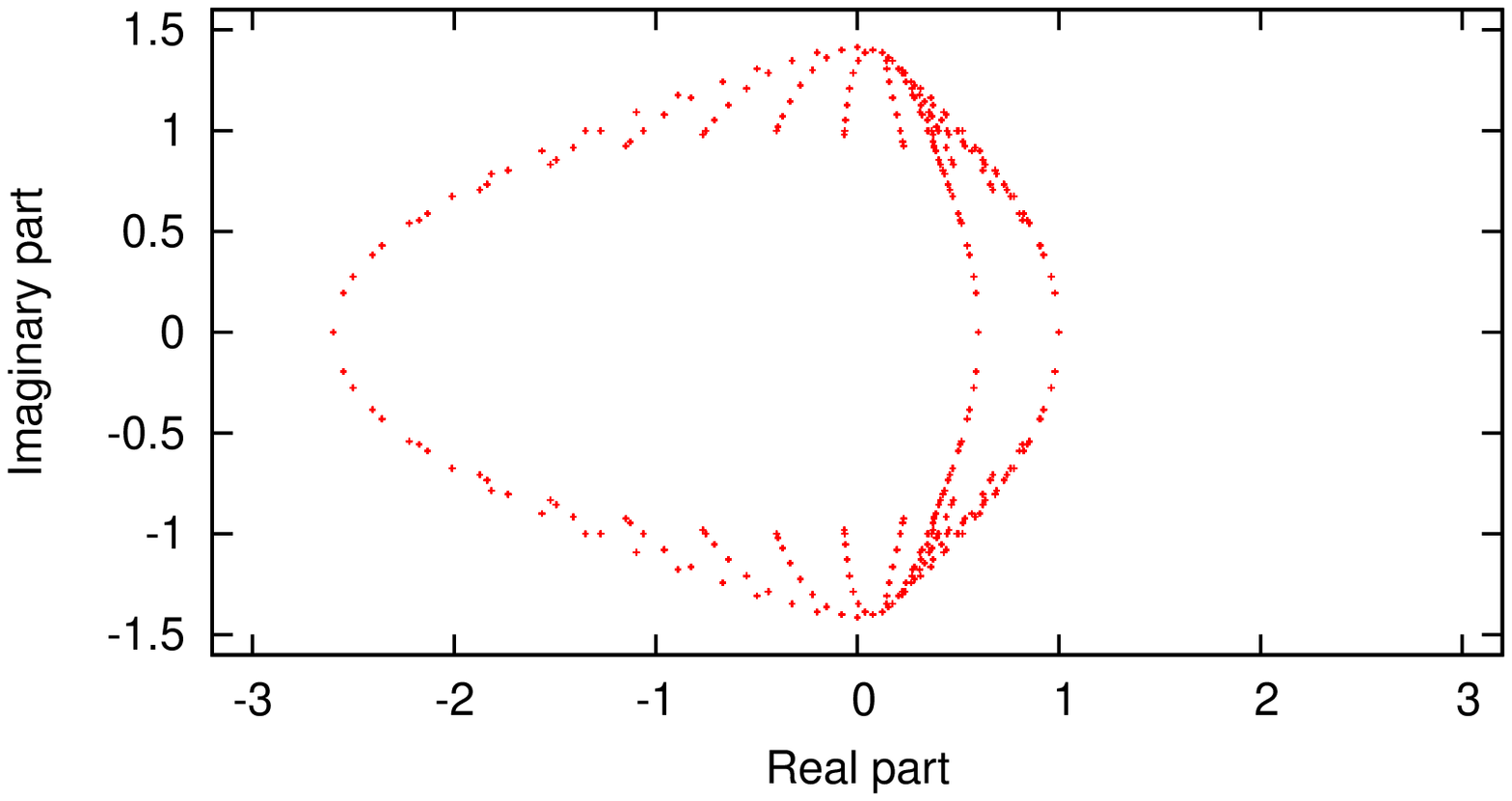}\\ 
  \vspace{-4em} \hspace{0.3em}
  \includegraphics[width=13em,angle=-90]{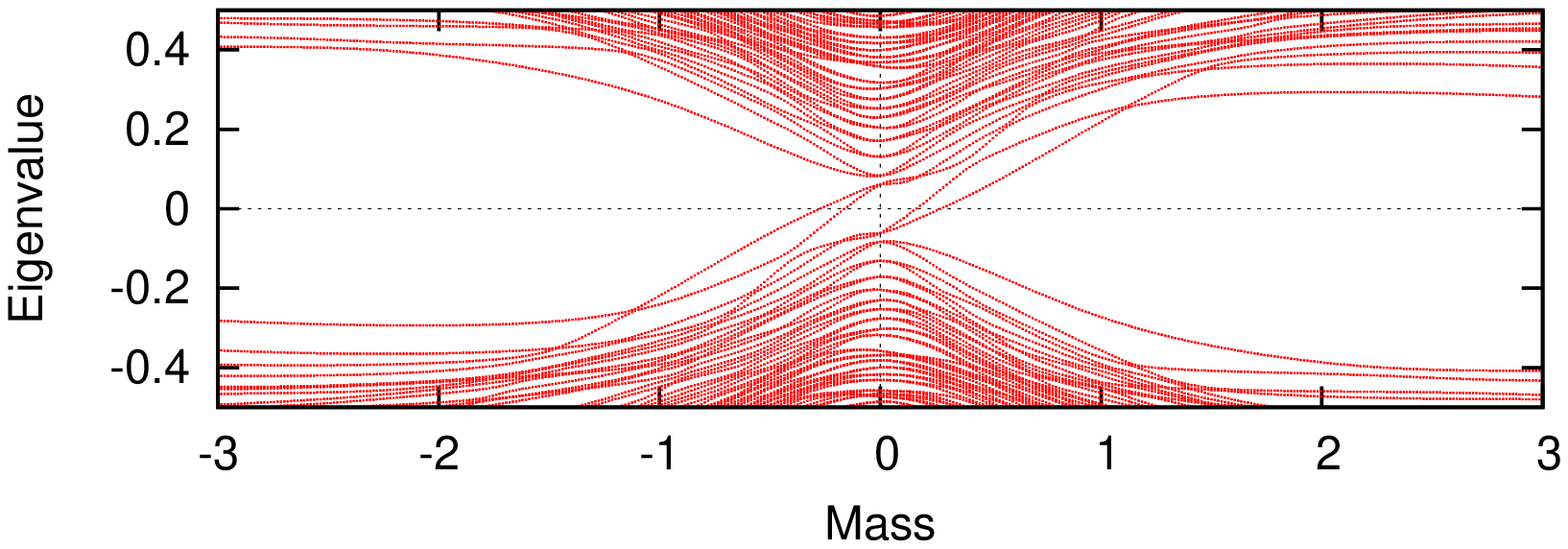}
  \includegraphics[width=13em,angle=-90]{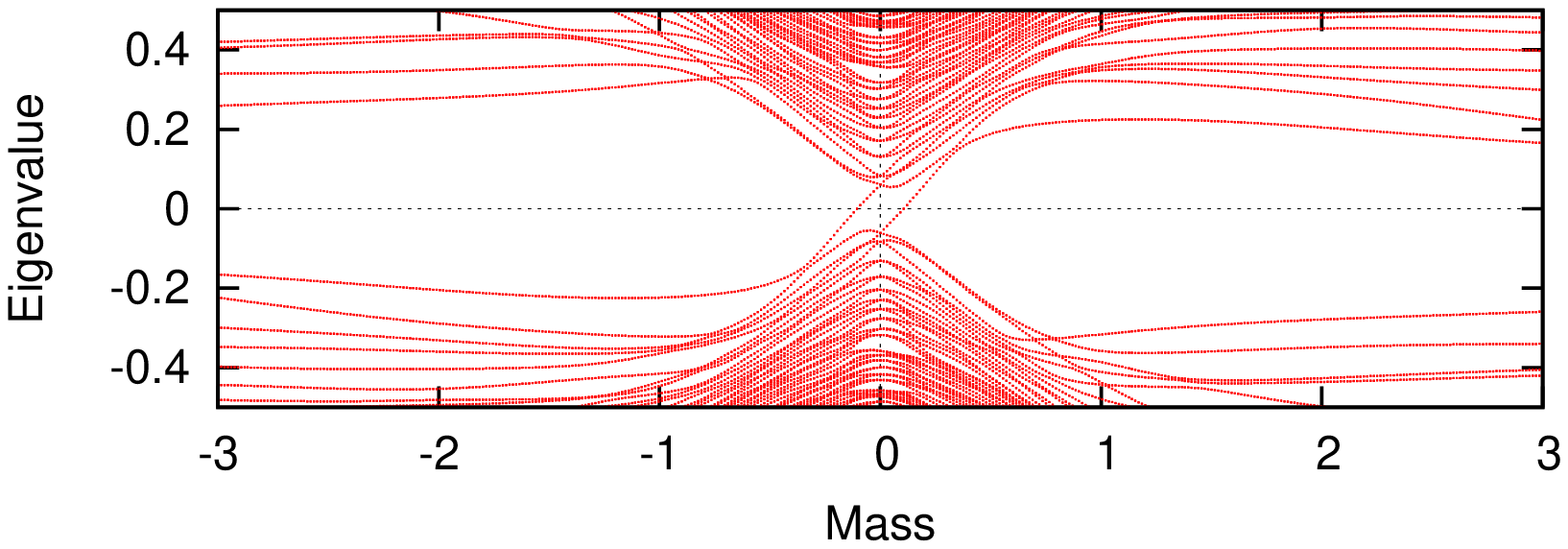}
  \vspace{-4em}
 \end{center}
 \caption{Dirac spectra and the corresponding spectral flows of the
 naive Hermitean operator with $M_{\rm naive}(c)$ for (left) $c=0.2$ and
 (right) $c=0.8$.}
 \label{sf_single}
\end{figure}

We then investigate how to reduce the number of massless modes of the
overlap operator with the naive kernel.
In the case of the $d=2$ naive fermion with the flavored-mass
term (\ref{flv_mass_n1}), there are two degenerate species with negative mass,
which leads to two overlap massless modes.
We lift this degeneracy by adding other kinds of flavored-mass 
terms in (\ref{flv_mass_n2}) and (\ref{flv_mass_n3}).
To preserve the rotational symmetry
we have to consider the following combination,
\begin{equation}
 M_{\rm naive}(c) = M_{\tau_3\otimes\tau_3} + c
  \left(M_{\tau_3\otimes\id} + M_{\id\otimes\tau_3}\right)
  \label{flv_mass_comb}
\end{equation}
with $m_{\tau_3\otimes\tau_3} = m_{\tau_3\otimes\id} = m_{\id\otimes\tau_3}$
in (\ref{flv_mass_n1})(\ref{flv_mass_n2})(\ref{flv_mass_n3}).
Fig.~\ref{sf_single} shows the Dirac spectra and the corresponding
spectral flows of the Hermitean operator built with this modified mass
term (\ref{flv_mass_comb}) for the two $c$ cases.
As seen from them, we change the number of negative-mass modes
in the original Dirac operator by choosing the parameter $c$.
Since such negative-mass modes are converted into massless modes in 
the overlap formulation, the result in Fig.~\ref{sf_single} indicates a single-flavor 
overlap fermion is obtained from the naive fermion kernel. 
The generalization to the $d=4$ is straightforward.
In terms of flavored-mass terms the difference between the original
overlap and our overlap fermions is just a choice of the flavored mass term.
Now we have varieties of overlap fermions including the staggered one and ours, 
which will be studied further.

\section{Summary}\label{sec:summary}

In this paper we have shown how the index theorem is realized in naive
and minimally doubled fermions by considering the spectral flow of the
Hermitean version of Dirac operators. 
The key is to make use of a point splitting for flavored mass terms. 
We also presented a new version of overlap fermions composed from the
naive fermion kernel, which is single-flavored and maintains the
hypercubic symmetry essential for a good continuum limit.

\end{document}